\documentstyle[11pt,psfig]{article}

\begin{document}
\begin{titlepage}
\begin{flushright}
\end{flushright}

\begin{center}
{\Large \bf Choice of Integrator in the Hybrid Monte Carlo Algorithm}

\vspace{1cm}
Tetsuya Takaishi

{ \it Hiroshima University of Economics \\
Hiroshima 731-0192, JAPAN}

\end{center}
\vspace{1cm}

\abstract{
We study efficiency of higher order integrator schemes 
for the hybrid Monte Carlo (HMC) algorithm.
Numerical tests are performed for Quantum Chromo Dynamics (QCD) with two flavors of Wilson fermions.
We compare 2nd, 4th and 6th order integrators at various quark masses.
The performance depends on both volume and  quark mass.
On currently accessible large lattices ( $V \sim 24^4$ ),
higher order integrators can be more efficient than the 2nd order one
only in heavy quark region, $m_q a > 0.3$.
Thus we conclude that for most full QCD simulations, 
except for heavy quark case,
the usual 2nd order integrator
is the best choice. 
}

\end{titlepage}

\section{Introduction}
Inclusion of dynamical fermions is one of major difficulties in lattice QCD
simulations 
since eventually one finds that the simulations require huge computational time.
The standard algorithm for full QCD simulations
is the hybrid Monte Carlo (HMC) algorithm\cite{HMCA}.
While the basic idea of the HMC is a combination of molecular dynamics ( MD )
and Metropolis accept/reject steps,
performance of its algorithm depends on tuning:
matrix solver, parameter tuning, integration scheme etc.
The matrix solver appears in the fermionic force calculations and
takes the dominant time in the HMC simulations.
The choice of an efficient matrix solver is an important subject 
to reduce CPU time \cite{SOLVER}. 
Parameters ( $\beta$ and $\kappa$ etc ) of a MD Hamiltonian
can be tuned so that the Metropolis acceptance rate increases \cite{Gupta_t}.

One may choose any integrator for the MD step provided that
the following two conditions are satisfied:
\begin{itemize}
\item area preserving
\item time reversibility
\end{itemize}

Usually the ( 2nd order ) leapfrog integrator is used for the HMC.
The leading integration errors are ${\cal O}(\Delta t^3)$, where
$\Delta t$ is the step size of an elementary MD step.
Due to these errors the Hamiltonian is not conserved. 
Let $\Delta H$ be an energy violation at the end of a MD trajectory.
To achieve the correct equilibrium
a new configuration should be accepted by a global Metropolis test
with a probability:
\begin{equation}
P \propto \min(1, \exp(-\Delta H)).
\end{equation}

In order to have a high acceptance we may consider
a more accurate integration scheme to reduce $\Delta H$.
The multiple time scale method \cite{SW} which removes 
the dominant errors from the gauge part worked well.
An idea \cite{Forcrand} which controls
the integration errors with an adaptive step size was also explored.
However no practical gain appeared for QCD case \cite{adaptive}.
One may employ a higher order integrator which has higher order 
integration errors in $\Delta t$. 
In general higher order integrators need more arithmetic operations
than the 2nd order one.
Therefore it is non-trivial whether  the higher order integrators serve
as an efficient speed-up source to the HMC.
For QCD  the  4th order one was studied on a $4^4$ lattice \cite{SW},
and was found not to be  efficient enough on such a small lattice.

When one compares higher order integrators, 
volume dependence must be considered.
The average acceptance of an {\bf n-th} order integrator is given by
$\sim{\rm erfc}(cV^{1/2}\Delta t^n)$\footnote{See Sec.4.}, 
where $V$ is volume of the system considered and $c$ is a constant.
To keep a constant acceptance, $\Delta t$ should scale $\sim 1/V^{1/2n}$.
This scaling behaviour suggests that the higher order one will be 
efficient for a lattice bigger than a certain size.
However we do not know the value of this lattice size.
In this study we perform HMC simulations with 2nd, 4th and 6th order 
integrators and clarify which integrator is efficient for a given lattice.
It now becomes feasible to perform a simulation on
rather big lattices as $24^3\times 40-48$ lattices\cite{FULL}.
So it is worthwhile to study whether, on such lattices,  the higher order 
integrators are more efficient than the standard leapfrog ( 2nd ) one.

In Sec.2 we describe the lattice QCD action used in our HMC simulations.
In Sec.3 we describe the higher order integrators which we use.
In Sec.4 we discuss the optimal efficiency, acceptance and step size of the HMC. 
In Sec.5 we give a criterion to compare various integrators.
In Sec.6 we present our numerical results.
Finally  we summarize our results in Sec.7.

\section{Lattice QCD action}

We use the standard plaquette gauge action  and
two flavors Wilson fermion action \cite{WILSON}.
The partition function is given by
\begin{equation}
Z=\int {\cal D}U \det[M(U)^{\dagger} M(U)] \exp(-S_g),
\label{PART}
\end{equation}
where $U$ stands for SU(3) link variables and
\begin{equation}
S_g = \frac{\beta}{3} \sum_{U_p} Tr(1-U_p)
\end{equation}
where $U_p$ stands for the plaquette and $\beta$ is the gauge coupling,
and the Wilson fermion matrix $M(U)$ is given by
\begin{equation}
M_{ij}(U)=\delta_{i,j}+\kappa \sum_{\mu}[(\gamma_\mu-1)U_{i,\mu}\delta_{i,j+\mu}
-(\gamma_\mu+1)U^{\dagger}_{i-\mu,\mu}\delta_{i,j+\mu}]
\label{MATRIX}
\end{equation}
where $\kappa$ is the hopping parameter.

The expectation value of some operator $O(U)$ is given by
\begin{equation}
<O>=\int {\cal D}U O(U) \det[M(U)^{\dagger} M(U)] \exp(-S_g)/Z.
\end{equation}

Using pseudofermion fields $\phi$ 
we replace  the determinant in Eq.(\ref{PART}) 
with a path-integral as
\begin{equation}
Z=\int {\cal D}U {\cal D}\phi^{*} {\cal D}\phi 
\exp[-S_g - \phi^\dagger (M(U)^\dagger M(U))^{-1}\phi].
\end{equation}

Introducing momenta $p$ conjugate to link variables
we define the Hamiltonian used in the HMC as
\begin{equation}
H=\frac{1}{2}p^2 +S_g +  \phi^\dagger (M(U)^\dagger M(U))^{-1}\phi
\end{equation}
and the partition function will be
\begin{equation}
Z=\int {\cal D}U  {\cal D}\phi^{*} {\cal D}\phi {\cal D}p \exp(-H),
\end{equation}
which gives the same expectation values as that from Eq.(\ref{PART}).

\section{Higher order integrators}
In this section we describe higher order integrator scheme which we use for the
present study.
Let $H$ be a classical Hamiltonian,
\begin{equation}
H=\frac{1}{2} p^2 + S(q)
\end{equation}
where $q=(q_1,q_2,...)$ and  $p=(p_1,p_2,...)$ are
coordinate variables and conjugate momenta respectively,
and $S(q)$ represents a potential term of the system.
For simplicity we use scaler variables $p$ and $q$.
The same discussion applies for QCD case where SU(3) link variables are used.

In the MD step we solve Hamilton's equations,
\begin{equation}
\frac{dq_i}{dt} = \frac{\partial H}{\partial p_i}
\end{equation}
\begin{equation}
\frac{dp_i}{dt} = -\frac{\partial H}{\partial q_i},
\end{equation}
approximately by an appropriate integrator.
In general these equations are not solvable analytically.
Let $T_{MD}(\Delta t)$ be an elementary MD step
with a time interval ( step size ) $\Delta t$,
which evolves $(p,q)$ to $(p^{\prime},q^{\prime})$:
\begin{equation}
T_{MD}(\Delta t):(p,q) \longrightarrow (p^{\prime},q^{\prime}).
\end{equation}
Requirements of the HMC to the integrator 
$T_{MD}(\Delta t)$ are (a) time reversible:
\begin{equation}
T_{MD}(-\Delta t):(p^{\prime},q^{\prime})  \longrightarrow (p,q)
\end{equation}
and (b) area preserving:
\begin{equation}
dpdq=dp^{\prime}dq^{\prime}
\end{equation}
i.e. invariance of the measure.
All integrators having the above requirements can be used for the HMC.
The simplest integrator is the 2nd order leapfrog method
which has been commonly used in the HMC of the
current full QCD simulations.
The 2nd order leapfrog scheme is explicitly written as

\begin{equation}
\left\{
\begin{array}{ll}
q(t+\frac{\Delta t}{2}) & =  q(t)+\frac{\Delta t}{2} p(t) \\
p(t+\Delta t) & =  p(t) -\Delta t \frac{\partial S(q(t+\frac{\Delta
t}{2}))}{\partial q} \\
q(t+\Delta t)& = q(t+\frac{\Delta t}{2}) +\frac{\Delta t}{2} p(t+\Delta t).
\end{array}
\right.
\end{equation}
While we start the integrator with variables $q$,
alternatively we can use momenta $p$ for the starting variables.
This 2nd order leapfrog integrator causes ${\cal O}(\Delta t^3)$ integration
error.

In order to construct a class of higher order integrators,
it is convenient to use the Lie algebraic formalism
\cite{SW,HOHMC,Yoshida,SYMP}.
The Hamilton's equation is written as
\begin{equation}
\frac{df}{dt} = \{f,H\}
\end{equation}
where $f=p$ or $q$, and \{,\} stands for the Poisson bracket, i.e.
\begin{equation}
\{f,H\}=\sum_i (\frac{\partial f}{\partial q_i} \frac{\partial H}{\partial
p_i}
- \frac{\partial H}{\partial q_i} \frac{\partial f}{\partial p_i}).
\end{equation}
Defining the linear ( Lie ) operator $L(H)$ as
\begin{equation}
L(H) f = \{f,H\},
\end{equation}
we have the formal solution of the Hamilton's equation:
\begin{equation}
f(t+\Delta t)=exp(\Delta t L(H) ) f(t).
\label{exp}
\end{equation}
Since $L(\cdot)$ is a linear operator, we have
\begin{eqnarray}
L(H) & = & L( \frac{1}{2}p^2) + L(S(q)) \\
     & = &  T + V
\end{eqnarray}
where $T\equiv L( \frac{1}{2}p^2)$ and $V\equiv L(S(q))$
stand for a kinetic and potential terms respectively.
Using the Lie algebraic formalism,
one finds that
the 2nd leapfrog integrator corresponds to a decomposition of
the exponential in Eq.(\ref{exp}) as
\begin{eqnarray}
f(t+\Delta t)& = &\exp(\Delta t (T+V) ) f(t) \nonumber \\
             & = & \{exp(\frac{1}{2}\Delta t T )exp(\Delta t V)
exp(\frac{1}{2}\Delta t T ) + {\cal O}(\Delta t^3)\} f(t).
\end{eqnarray}
Note that $T$ and $V$ do not commute with each other and ${\cal O}(\Delta t^3)$ decomposition errors appear.
An important observation here is that
the order of the decomposition error
coincides with that of the integration error.

An arbitrary order integrator can be found by decomposing
the exponential with the desired order.
In general, the exponential is decomposed as
\begin{equation}
exp(\Delta t (T+V) ) =  \prod_i exp (c_i \Delta t T) exp (d_i \Delta t V) +
{\cal O}(\Delta t^{n+1})
\end{equation}
where $c_i$ and $d_i$ are determined so that the decomposition is correct up
to ${\cal O}(\Delta t^{n})$.
It is not obvious how to obtain such $c_i$ and $d_i$ in any order.
Fortunately higher {\it even}-order integrators are known to be constructed
from a combination of lower order integrators\cite{HOHMC,Yoshida,Suzuki}.
Let us call $G_{2nd}(\Delta t)$ the 2nd order decomposition
( or integrator),
\begin{equation}
G_{2nd}(\Delta t) \equiv exp(\frac{1}{2}\Delta tT )exp(\Delta t V)
exp(\frac{1}{2}\Delta t T ).
\end{equation}

The 4th order integrator is given by a product of $three$ 2nd order
integrators\cite{4th,HOHMC,Yoshida,Suzuki},
\begin{equation}
G_{4th}(\Delta t) = G_{2nd}(a_1 \Delta t) G_{2nd}(a_2 \Delta t) G_{2nd}(a_1
\Delta t)
\end{equation}
where the coefficients $a_i$ are given by
\begin{equation}
a_1 = \frac{1}{2-2^{1/3}} ,
\end{equation}
\begin{equation}
a_2 = - \frac{2^{1/3}}{2-2^{1/3}}.
\end{equation}
This construction scheme is easily generalized for an arbitrary
even-order one{\cite{HOHMC,Yoshida,Suzuki}.
(2k+2)-th order integrator is given recursively by
\begin{equation}
G_{2k+2}(\Delta t) = G_{2k}(b_1 \Delta t) G_{2k}(b_2 \Delta t) G_{2k}(b_1
\Delta t),
\label{rec}
\end{equation}
where the coefficient $b_i$ are
\begin{equation}
b_1 = \frac{1}{2-2^{1/(2k+1)}}
\end{equation}
\begin{equation}
b_2 = - \frac{2^{1/(2k+1)}}{2-2^{1/(2k+1)}}.
\end{equation}

While one can find an arbitrary higher even-order integrator with Eq.({\ref{rec}),
the number of elementary steps ( 2nd order integrator ) grows with the order
of the integrator as follows:
\begin{equation}
\# \mbox{ of  2nd order integrator} = \left\{
\begin{array}{ll}
1 & 2nd \\
3 & 4th \\
9 & 6th \\
\vdots & \vdots \\
3^{n/2-1} & nth.
\end{array}
\right.
\label{cost}
\end{equation}
A bottleneck of the HMC for QCD is the force calculation $\partial S/\partial
q$
which needs a large amount of computational time
devoted to a matrix solver.
Except for some small overhead, the computational cost of the HMC is
proportional to the number of the force calculations.
The 2nd order integrator contains one force calculation.
Therefore the computational cost of the higher order
algorithm can be counted by Eq.(\ref{cost}),
which  indicates that the cost grows rapidly with the order.

If one can find a higher order integrator consisting of 
fewer force calculations it may be useful for HMC.
In Ref\cite{Yoshida}, such a higher order scheme is found numerically.
In this study we also use the 6th order integrators of Ref\cite{Yoshida}
which consist of $seven$ 2nd order integrators instead of $nine$ as in
Eq.(\ref{cost}).
The 6th order integrators of Ref\cite{Yoshida} are written as
\begin{eqnarray}
G_{6th}(\Delta t)&=&G_{2nd}(w_3 \Delta t) G_{2nd}(w_2 \Delta t) G_{2nd}(w_1
\Delta t)G_{2nd}(w_0 \Delta t) \nonumber \\
& & \times G_{2nd}(w_1 \Delta t) G_{2nd}(w_2 \Delta t) G_{2nd}(w_3 \Delta
t),
\label{6th}
\end{eqnarray}
where values of $(w_0,w_1,w_2,w_3)$ are listed in Table 1.

Note that all the higher even-order integrators described here satisfy the time
reversible and area preserving conditions
since those are a product of the 2nd order integrators having the area
preserving condition
and are constructed in a symmetric way ( $G(\Delta t)G(-\Delta t)=1$ ) which
yield the time reversible condition.

\begin{table}[h]
\begin{tabular}{c|ccc}  \hline
    & Y1 & Y2& Y3 \\ \hline
$w_1$ & -0.117767998417887e-1 & -0.2132285222000144e+1 &
0.152886228424922e-2 \\
$w_2$ & 0.235573213359357e+0  & 0.426068187079180e-2
  & -0.214403531630539e+1 \\
$w_3$ & 0.784513610477560e+0  & 0.143984816797678e+1   &
0.144778256239930e+1  \\ \hline
\end{tabular}
\caption{Parameter sets (Y1-Y3) of the 6th order integrators
by Yoshida {\cite{Yoshida}}.
$w_0$ is given by $w_0=1-w_1-w_2-w_3$.
}
\end{table}

\section{Optimal efficiency, acceptance and step size}

In this section we introduce an efficiency function which
characterizes the speed of algorithm and
derive formulae for the optimal acceptance and step size
which define the optimal efficiency. 

We define the efficiency function $E_{ff}$ by
a product of step size $\Delta t$ and acceptance $P_{acc}(\Delta t)$:
\begin{equation}
E_{ff}(\Delta t) = P_{acc}(\Delta t) \Delta  t.
\label{eff}
\end{equation}
High $E_{ff}$ results in fast Markov step when producing configurations. 
A particular length of trajectory does not affect $E_{ff}$
since the acceptance stays almost constant
for any trajectory length longer than a certain characteristic
length\cite{Accept}.
This is a feature of the symplectic type integrator \cite{SYMP}.
We fix the trajectory length to the unit length (=1)
which is sufficiently longer than the characteristic length for the present
study\footnote{Several simulations have been done 
with both trajectory lengths of
0.5 and 1.0. We found no significant change in the acceptance among them.}.

The acceptance decreases as $\Delta t$ increases.
In both limits of $\Delta t= 0$ and $\infty$,
$E_{ff}$ goes to zero.
We expect that $E_{ff}$ has one maximum at a certain $\Delta t$,
which we call optimal step size $\Delta t_{opt}$.

Using $\Delta t_{opt}$ we define the optimal acceptance $(P_{acc})_{opt}$:
\begin{equation}
(P_{acc})_{opt} = P_{acc}(\Delta t_{opt}),
\label{Popt}
\end{equation}
and the optimal efficiency $(E_{ff})_{opt}$:
\begin{eqnarray}
(E_{ff})_{opt} & = & \max(E_{ff}(\Delta t)) = E_{ff}(\Delta t_{opt}) \\
                & = & (P_{acc})_{opt} \Delta t_{opt}.
\end{eqnarray}

When we have $(E_{ff})_{opt}$ the maximum speed of the algorithm is achieved.
Comparison among various integrators will be done with this $(E_{ff})_{opt}$.
At this stage, however, it is not obvious how to
obtain $(E_{ff})_{opt}$ easily from Monte Carlo (MC) simulations.
In the following consideration, we show that one coefficient
governs $(E_{ff})_{opt}$ and it is easily obtainable  from a MC
simulation.

At large volumes, the average acceptance with an average energy difference
is evaluated as \cite{Accept}
\begin{eqnarray}
\langle P_{acc}\rangle& =& {\rm erfc}(\frac{1}{2}\langle \Delta
H\rangle^{1/2}).
\label{acc}
\end{eqnarray}
Although Eq.(\ref{acc}) is applicable for the whole range of
the acceptance, i.e. $1 \geq \ \langle P_{acc}\rangle  \geq 0$,
we are not interested in low acceptance with which the algorithm
may not be efficient.
Typically we may need $\langle P_{acc}\rangle >50\%$.
Instead of Eq.(\ref{acc}) we propose a simple exponential type formula:
\begin{equation}
\langle P_{acc}\rangle=\exp(-\frac{2}{\sqrt{\pi}}\langle \frac{1}{8}\Delta
H^2\rangle^{1/2}).
\label{Papprox}
\end{equation}
In the limit of $\langle \Delta H\rangle \rightarrow 0$,
Eq.(\ref{Papprox}) coincides with Eq.(\ref{acc}) ( See Appendix ).
Even if  $\langle \Delta H\rangle$ is not small enough 
(  $\langle \Delta H\rangle \simeq 1$ ),
from numerical tests 
we confirm that Eq.(\ref{Papprox}) is a good approximation to the exact value. 
Fig.1 shows comparison of  MC results and
values from Eq.(\ref{Papprox}).
Fig.2 shows a normalized error [(MC$-$Eq.(\ref{Papprox}))/Eq.(\ref{Papprox})]
as a function of $\langle \Delta H^2\rangle^{1/2}$.
From the comparison of MC results and Eq.(\ref{Papprox})
we notice that Eq.(\ref{Papprox})  
agrees quite well with the MC results 
within $5\%$ error up to $\langle \Delta H^2\rangle^{1/2}  \leq 3$,
which roughly corresponds to $\langle P_{acc}\rangle \geq 20\%$.
Therefore we adopt Eq.(\ref{Papprox})  as our formula for $\langle P_{acc}\rangle$.

Let us now discuss $\Delta t$ dependence of $\langle \Delta
H^2\rangle^{1/2}$
which appears in Eq.(\ref{Papprox}).
The 2nd order integrator causes ${\cal O}(\Delta t^3)$ integration errors 
after an elementary MD step.
From the discussion of Ref.\cite{Accept}, however, 
$\Delta H \sim \Delta t^2$ rather than $\sim \Delta t^3$.
This comes from the fact that $\Delta H$ does not grow with 
the trajectory length 
( = the number of elementary MD steps $\times \Delta t$ ). 
Using $\langle \Delta H\rangle \approx \frac{1}{2}\langle \Delta H^2\rangle$
which is expected from Creutz's equality\cite{Creutz} 
$\langle \exp(-\Delta H)\rangle=1$
at small $\Delta H$,
we obtain
\begin{equation}
\langle \Delta H\rangle \sim \langle \Delta H^2\rangle \sim V \Delta t^4
\end{equation}
where $V$ is the volume of the system.
When we apply the same discussion for the n-th order integrator
we have
\begin{equation}
\langle \Delta H^2\rangle \sim V \Delta t^{2n}.
\end{equation}
Thus,
\begin{equation}
\langle \Delta H^2\rangle^{1/2} \sim V^{1/2} \Delta t^n.
\label{dH}
\end{equation}

For our later use
we rewrite Eq.(\ref{dH}) as
\begin{equation}
\langle \Delta H^2\rangle^{1/2} = C_n V^{1/2} \Delta t^n +
{\cal O} (\Delta t^{n+1}),
\label{dH2}
\end{equation}
where $C_n$ is a Hamiltonian ( model ) dependent coefficient, which is not
known a priori.
We checked Eq.(\ref{dH2}) by MC simulations, as shown in Fig.3.
When $\Delta t$ is not too large, Eq.(\ref{dH2}) holds.

Here we comment on the 6th order integrator. In the MC simulations for
Fig.3, the standard construction scheme of
Eq.(\ref{rec}) was used.
We have other 6th order integrators defined by  Eq.(\ref{6th}) 
which have less computational cost.
These have the same power of $\Delta t$ to the leading term.
The proportional coefficient $C_n$ ( here $n=6$ ), however, can be different
for each 6th order integrator.
We numerically calculated $C_n$ of each integrator
using quenched QCD and quenched Schwinger ( QED$_2$ ) models.
Fig.4 shows $\langle \Delta H^2\rangle^{1/2}$ as a function of $\Delta t$, where
{\bf Stand} indicates the standard construction scheme of
Eq.(\ref{rec}).
Others are from Eq.(\ref{6th}).
The most efficient one is the integrator {\bf  Y1} in Table 1, of which $C_n$
is roughly a factor of ten smaller than others.
The results from others  are more or less same.
The same conclusion is also applied for the Schwinger model as shown in
Fig.5.
It seems that {\bf Y1} is always the most efficient one.
Thus in the following analysis we use  {\bf Y1}
as our 6th order integrator.

Using Eq.(\ref{dH2}) without higher order terms, we find a formula for
the average acceptance, instead of  Eq.(\ref{Papprox}), as
\begin{equation}
\langle P_{acc}\rangle = \exp(-\widetilde{C}_n V^{\frac{1}{2}} \Delta t^n),
\label{Accaprox}
\end{equation}
where $\widetilde{C}_n = C_n/\sqrt{2\pi}$.
Using this expression, we can easily obtain the optimal step size as
\begin{equation}
\Delta t_{opt} = \sqrt[n]{\frac{1}{n \widetilde{C}_n V^{\frac{1}{2}}}  }.
\label{DTopt}
\end{equation}
Substituting this result to Eq.(\ref{Popt}) we obtain the optimal acceptance as
\begin{eqnarray}
\langle P_{acc}\rangle_{opt} & = &   \exp(-\frac{1}{n}) \label{PAopt} \\
& =&  \left\{
\begin{array}{ll}
0.61 & \mbox{2nd} \\
0.78 & \mbox{4th} \\
0.85 & \mbox{6th}.
\end{array}
\label{PAopt2}
\right.
\end{eqnarray}
Finally from Eq.(\ref{DTopt}) and Eq.(\ref{PAopt})
we obtain the optimal efficiency of the n-th order integrator:
\begin{equation}
(E_{ff})_{opt}^{n-th}= \exp(-\frac{1}{n}) \sqrt[n]{\frac{1}{n
\widetilde{C}_n V^{\frac{1}{2}}}  },
\label{Eopt}
\end{equation}
and find that Eq.(\ref{Eopt}) is governed by only one unknown $\widetilde{C}_n$.

The result of Eq.(\ref{PAopt}) suggests that there exists an optimal
acceptance depending
only on the order of the integrator, not on the model.
The optimal acceptance increases with increase of the order of the
integrator.
We verify this feature by MC simulations.
First we show results of QCD for 2nd order one
and then that of quenched Schwinger model ( QED$_2$ )
for 2nd,4th and 6th order ones.
The Schwinger model is used to reduce the CPU time.

In Fig.6 results for the 2nd order integrator
from three parameter sets among  different $\beta=(5.3, 5.0, 0.0)$,
$\kappa=(0,0.2)$ and  volume $V=(6^4,4^4)$ are plotted.
For all the cases the optimal acceptance locates around a value 
between $60-70\%$ which is
in good agreement with the analytic estimate ( 61\% ) of  Eq.(\ref{PAopt2}).
The results with different lattice volume compare lattice size dependence on
$(E_{ff})_{opt}$.
Since QCD is a 4-dimensional model ( $V=L^4$ where $L$ is lattice size )
we find  $(E_{ff})_{opt} \sim 1/L$ for the 2nd order integrator.
The MC results from $4^4$ and $6^4$ lattices
with slightly different $\beta$ agree with this expectation,
i.e.  $(E_{ff})_{opt}[4^4]/(E_{ff})_{opt}[6^4]\sim 1.5$ ( See Fig.6 ).

Fig.7 compares  $(E_{ff})$ among 2nd, 4th and 6th integrators.
It is clearly seen that the optimal acceptance
increases with increase of the order, which is again in agreement with
Eq.(\ref{PAopt2}).

\section{Comparison of various integrators}

In this section we give a criterion to  compare various integrators.
Let us consider the {\bf n-th} and {\bf m-th} order integrators ($n>m$).
As seen in the previous section,
each integrator has the optimal efficiency given by Eq.(\ref{Eopt}).
In order to have a better performance for the n-th order integrator
than for the m-th one, the optimal efficiency of the n-th one
should be larger than that of the m-th one.
Furthermore we must consider the cost to perform the higher order one
since the higher order one needs more arithmetic operations.
Thus the following equation should be satisfied,
\begin{equation}
(E_{ff})^{n-th}_{opt}> k_{nm} (E_{ff})^{m-th}_{opt}
\label{comp}
\end{equation}
where $k_{nm}$ is a relative cost factor needed to implement the n-th order
integrator against the m-th one.

Substituting Eq.(\ref{Eopt}) to Eq.(\ref{comp}),
we obtain
\begin{equation}
\exp(-\frac{1}{n})\sqrt[n]{\frac{1}{n \widetilde{C}_n V^{\frac{1}{2}}}  } >
k_{nm} \exp(-\frac{1}{m})\sqrt[m]{\frac{1}{m \widetilde{C}_m
V^{\frac{1}{2}}}  }.
\label{comp2}
\end{equation}

Rewriting Eq.(\ref{comp2}),
we obtain an expression for volume size with which
the n-th order integrator performs better than the m-th order one,
\begin{equation}
V^{\frac{n}{2}-\frac{m}{2}} > (k_{nm}\exp(-\frac{1}{m}+\frac{1}{n}))^{n m}
\left(\frac{1}{m\widetilde{C}_m}\right)^{n} (n\widetilde{C}_n)^m.
\label{comp3}
\end{equation}

In the present study we compare (A): 2nd and 4th order integrators, and
(B): 4th and 6th order integrators.

(A): 4th order versus 2nd order

In this case $n=4$ and $m=2$. From Eq.(\ref{cost}) we find
the relative cost is $k_{4,2}=3$.
Substituting $k_{4,2}=3$ into Eq.(\ref{comp3}) we obtain
\begin{equation}
V^{\frac{1}{2}}  > (3\exp(-\frac{1}{4}))^4
\left(\frac{1}{\widetilde{C}_2}\right)^2 \widetilde{C}_4.
\label{CASE42}
\end{equation}

(B): 6th order versus 4th order

In this case  $n=6$ and $m=4$. Since we use the scheme of Eq.(\ref{6th}),
the relative cost $k_{6,4}$ is $7/3$. Thus we obtain
\begin{equation}
V^{\frac{1}{2}} > \left(\frac{7}{3}\exp(-\frac{1}{12})\right)^{12}
\left(\frac{1}{4\widetilde{C}_4}\right)^3
\left(6\widetilde{C}_6\right)^2.
\label{CASE64}
\end{equation}

\section{Lattice size for higher order integrator}

Now we come to the stage of determination of lattice sizes
which are suitable for the higher order integrators.
Eqs.(\ref{CASE42})-(\ref{CASE64}) determine
regions where the higher order integrators perform better than the lower one.
In practice we solve Eqs.(\ref{CASE42})-(\ref{CASE64})
equating both sides of the equations.
The solutions ( in lattice size ) form  a boundary
which separates two regions:
higher order and  lower order preferred regions.
Unknown $\widetilde{C}_n$ should be obtained from numerical simulations.
We choose a small enough $\Delta t$ and
compute $\langle \Delta H^2\rangle^{1/2}$.
Applying Eq.(\ref{dH2}) for $\langle \Delta H^2\rangle^{1/2}$
we extract $C_n(=\sqrt{2\pi}\widetilde{C}_n)$.

First we study quenched QCD where $C_n$
are determined as a function of $\beta$,
and then go to full QCD with two flavors of Wilson fermions  
where $C_n$ are a function of $\kappa$.

\subsection{Quenched QCD}
Numerical simulations were performed on a $4^4$ lattice.
Fig.8 shows $C_n$ as a function of $\beta$.
We also used an $8^4$ lattice at several $\beta$
to check the volume dependence appearing in Eq.(\ref{dH2}).
The values of $C_n$ obtained from both lattices were same,
which suggests that we can get reliable values of $C_n$ on the lattice of this size. 

Fig.9 shows results of the boundaries determined from Eqs.(\ref{CASE42})-(\ref{CASE64}).
The circle symbols form a boundary which separates the 2nd order
preferred region ( lower region ) and the 4th order preferred one ( upper
region ).
Similarly the squares separate
the 4th order one ( lower ) and the 6th order one ( upper ).

The boundary between the 2nd order one and the 4th order one ({\bf B2-4})
increases as $\beta$ increases and
reaches a plateau at $\beta\sim3.0$. The corresponding lattice size is
about $\sim10^4$.
On the other hand the boundary between the 4th and the 6th one ({\bf B4-6})
decreases with $\beta$. At $\beta\sim5.0$ the lattice size on {\bf B4-6} is
about $20^4$.
At $\beta\sim 5.0$, the 4th order integrator becomes efficient for $L>10$ and
the 6th order one for $L>20$.

\subsection{Full QCD}
We use a model with two flavors of Wilson fermions.
To consider fermion dynamics only we take $\beta=0$ and
simulate the model with  varying $\kappa$.
An advantage of taking $\beta=0$ is that
the critical kappa is known analytically, i.e. $\kappa_c=0.25$.
Using the critical kappa
the quark mass at $\beta=0$ is
defined by $\widetilde{m}_q =m_q a =\ln(1+\frac{1}{2}(1/\kappa -1/\kappa_c))$\footnote{The alternative definition $m_q a=\frac{1}{2}(1/\kappa -1/\kappa_c)$ 
gives similar results}.

Fig.10 shows $C_n$ as a function of quark mass.
The simulations were done on a $4^4$ lattice.
We see that $C_n$ behaves as a power of quark mass, i.e. $C_n \propto
\widetilde{m}_q^{-\alpha}$.
Using  three data at small $\widetilde{m}_q( \kappa=0.215,0.225,0.230)$,
$\alpha$ are estimated to be $\alpha \sim 1.55(8), 4.58(11), 7.65(11)$ for
2nd, 4th and 6th order respectively.
For the 2nd order, this result is consistent with that of
the staggered fermion\cite{Accept}: $\alpha\approx 1.5$.

Fig.11 shows results of {\bf B2-4} and {\bf B4-6}.
Both  {\bf B2-4} and {\bf B4-6} increase with decreasing quark mass.
We estimate quark mass dependence of the boundaries as:
{\bf B2-4} $\propto C_4^{1/2}/C_2\sim \widetilde{m}_q^{-0.74}$ 
and  {\bf B4-6} $\propto C_6/C_4^{3/2} \sim \widetilde{m}_q^{-0.73}$.
For both, values of the power are negative,
which means that at small quark masses, the lattice size needed to have a gain 
with the higher order integrator increases. 
If we stay at $L<24(40)$ which is a lattice size available for 
the current ( or near-future ) full QCD simulations,
the 4th order integrator can be efficient for $\widetilde{m}_q>0.3(0.1)$.
The chiral limit ( $m_q \rightarrow 0$ ) is the primary interesting
case in full QCD simulations.
Therefore our results show that the standard leapfrog ( 2nd order ) integrator
is the best one for most full QCD simulations except for heavy quarks. 

For comparison we also give results of the Schwinger model 
with staggered quarks at $\beta=0.0$.
Fig.12 shows $C_n$ as a function of staggered quark mass.
The behaviour of $C_n$ is similar to that of the full QCD case, 
i.e. $C_n \propto\widetilde{m}_q^{-\alpha}$ 
: $\alpha$ is estimated to be 1.36(9), 3.36(27) and 5.41(23) for
2nd, 4th and 6th integrators respectively.
The estimation is based on the data at small quark masses.
Fig.13 shows {\bf B2-4} and {\bf B4-6}. Here note that 
$V=L^2$ since the Schwinger model is a 2 dimensional model. 

Although we considered $\beta=0.0$ case only, at finite $\beta$  
we expect the similar diagram as in Fig.11 for small $m_q a$.
The reason is the following.
The fermionic force is expected to become dominant over 
the gauge force at small $m_q a$.
In such a case, the integration error from the fermionic part also
becomes dominant at small $m_q a$.
As seen in Fig.8 all $C_n$ of quenched QCD at $\beta \sim 5.0$ are less than 10.
On the other hand all $C_n$ at $m_q a < 0.2$ are bigger than 100.
Therefore the naive expectation is 
that the dominant contribution to $\langle \Delta H^2\rangle^{1/2}$
at small $m_q a$ comes from the fermionic part
and values of $C_n$ at small $m_q a$
may not drastically change from that of $C_n$ at $\beta=0.0$.
No change of $C_n$ results in no change of 
the boundaries ( {\bf B2-4}  and {\bf B4-6} ).

We might also expect that the multiple time scale method \cite{SW}
does not work for small $m_q a$
since it integrates finely only the gauge part
and the dominant error from the fermionic part remains big
at small quark masses.

To justify the above  expectations, we calculate $C_n$ at various $\beta$ and
illustrate how the dominant  contribution to $\langle \Delta H^2\rangle^{1/2}$
comes from the fermionic part.  
For this purpose we choose the Schwinger model with staggered fermions, 
which requires less CPU time.
Fig.14 shows $C_2$ versus $m_qa$ at $\beta=0.0$, 0.5 and 1.0.
At large $m_qa$, the fermionic contributions are expected to be small and,
in the limit of $m_qa\rightarrow \infty$ the values of $C_n$ go to
those of quenched ones.
In contrast to the situation at large $m_qa$, 
as $m_qa$  goes to small quark masses,  
values of $C_2$ at finite $\beta$ become similar 
to those of $C_2$ at $\beta=0.0$, which shows that the fermionic contribution 
becomes dominant and the values of $\beta$ are less important. 
We observed similar results for the higher order integrators. 
Therefore we deduce the similar result to that at $\beta=0.0$.

\section{Summary}
We have investigated higher order integrators for HMC
with a criterion which compares among various integrators. 
The criterion is governed by one unknown parameter which is
easily obtainable from a MC simulation.
We have made comparison with quenched and full QCD models.

For quenched QCD the 4th order integrator performs better
than the 2nd one for a lattice with size $L>10$ at $\beta \sim 5.0$.
Of course usually the HMC is not used for quenched QCD simulations.
If one uses a complicated action which is not implemented effectively
with a local update algorithm
the HMC with a higher order integrator could be
an efficient algorithm for its simulation.

For full QCD the higher order integrators can be efficient 
only for large $m_q a$.  
For instance, on a currently accessible big lattice ( $L\sim 24$ )
the 4th order one performs better than the 2nd order one
only  for $m_q a > 0.3$, which is out of interest for most full QCD
simulations. Thus the 2nd order one is the best one
for the current full QCD simulations.

The higher order integrators are turned out to be 
uninteresting at low $\beta$ ( Fig.9 ) and small $m_q a$ ( Fig.11 ).
This may be due to the fact that QCD becomes more perturbative (i.e. 
close to a integrable system)  at high $\beta$ and large $m_q a$. 
The Hamilton's equations can be effectively integrated with 
higher order integrators in perturbative region.  
On the other hand, at low $\beta$ and small $m_q a$ 
(in non-perturbative region), 
the higher order integrators may not be efficient enough.

The optimal acceptance strongly depends on the order of
the integrator. An interesting case is the 2nd order one,
where the optimal acceptance is measured to be around 60-70\%
( analytically estimated to be 61\% ).
This indicates that a very high acceptance like 80-90\% or more
is not required for the HMC with the 2nd order integrator.
We suggest to take an acceptance around 60-70\% for
2nd order HMC simulations of any model.

We have not considered finite temperature case.
Ref.\cite{Accept} found that quark mass dependence of 
the coefficient ( here $\sim C_2$ )
is very weak in the finite temperature phase.
If this is also true for the higher order integrators
the boundaries may not increase rapidly with decreasing quark mass
as fast as in the zero temperature case.

\vspace{1cm}
{\large \bf ACKNOWLEDGMENTS}

The author would like to thank Ph. de Forcrand for 
valuable comments and helpful discussions.
He is also grateful to A.Nakamura and O.Miyamura for discussions. 
This work was supported in part by Hiroshima University of Economics,
and by the Grant in Aid for Scientific Research by the Ministry 
of Education (No.11740159).

\vspace{1cm}
\appendix{\large \bf APPENDIX}

When the argument $x$ is small, the error function erfc($x$) is approximated as
follows.
\begin{eqnarray}
{\rm erfc}(x) & = & 1-{\rm erf}(x) \nonumber \\
        & = & 1-\frac{2}{\sqrt{\pi}}\int^x_0 dt e^{-t^2} \nonumber \\
        & = & 1-\frac{2}{\sqrt{\pi}}(x+x^2+...) \nonumber \\
        & \approx & \exp(-\frac{2}{\sqrt{\pi}}x).
\end{eqnarray}
Thus,
using $\langle\Delta H\rangle \approx \frac{1}{2}\langle\Delta H^2\rangle$
at small $\langle\Delta H\rangle$,
Eq.(\ref{acc}) is approximated as
\begin{eqnarray}
\langle P_{acc}\rangle & =& {
\rm erfc}(\frac{1}{2}\langle\Delta H\rangle^{1/2}) \\
         & \approx & {\rm erfc}(\langle\frac{1}{8}\Delta H^2\rangle^{1/2})
\\
         & \approx & \exp(-\frac{1}{\sqrt{2\pi}}\langle\Delta
H^2\rangle^{1/2}).
\end{eqnarray}


\newpage

\begin{figure}
\centerline{\psfig{figure=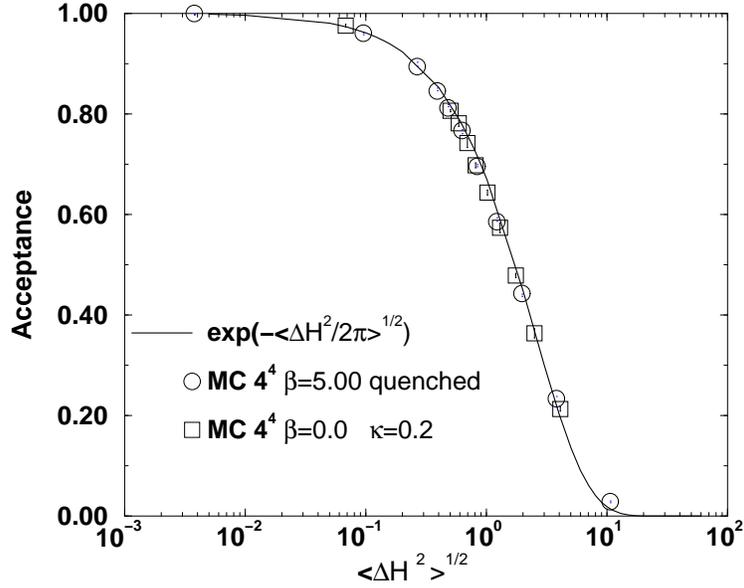,height=8cm}}
\caption{
Comparison of the average acceptance between 
$\exp(-\langle \Delta H^2 \rangle^{1/2}/ \sqrt[]{2 \pi})$
and MC results as a function of $\langle\Delta H^2\rangle^{1/2}$.  
}
\end{figure}

\begin{figure}
\centerline{\psfig{figure=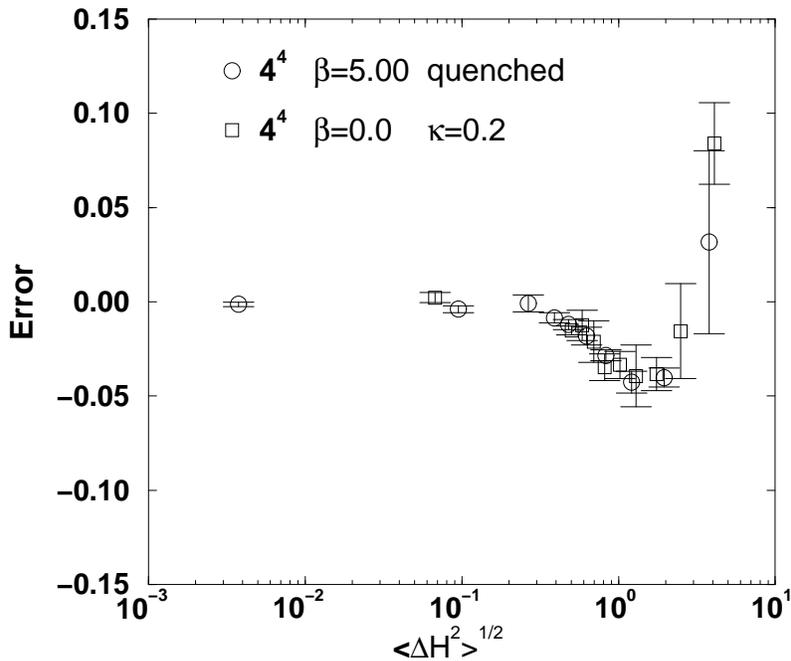,height=9cm}}
\caption{
Difference between $\exp(-\langle\Delta H^2\rangle^{1/2}/\sqrt{2\pi})$
and MC results.
Error is defined by (MC data - $x$)/$x$, 
where $x=\exp(-\langle\Delta H^2\rangle^{1/2}/\sqrt{2\pi})$. 
}
\end{figure}

\begin{figure}
\centerline{\psfig{figure=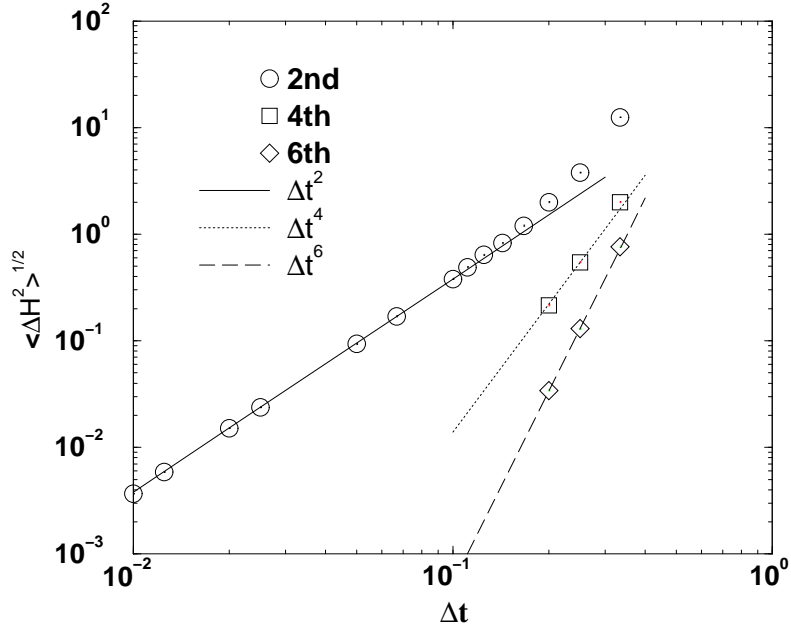,height=8.5cm}}
\caption{
$\langle\Delta H^2\rangle^{1/2}$ for the 2nd, 4th and 6th order integrators
as a function of $\Delta t$. The simulations ( quenched QCD ) were done
on a $4^4$ lattice at $\beta=5.0$. Three lines in the figure are shown
to guide the eye.
}
\end{figure}

\begin{figure}
\centerline{\psfig{figure=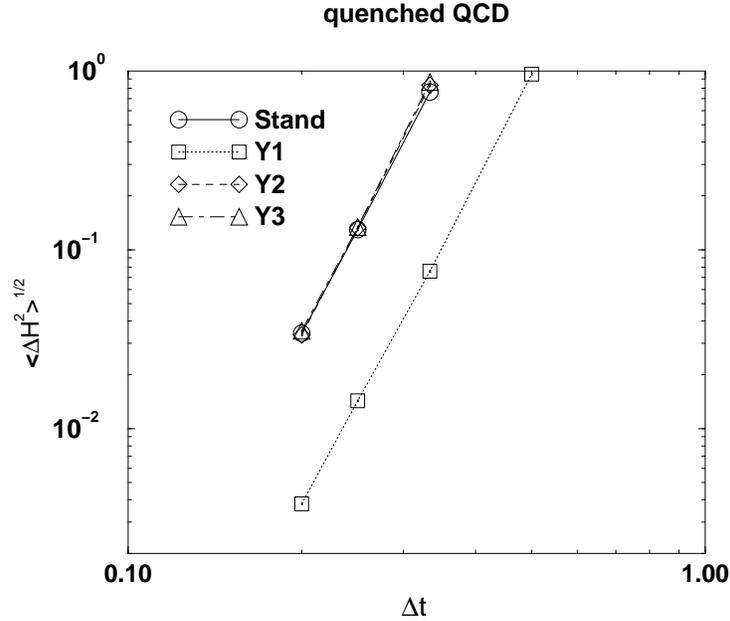,height=8.5cm}}
\caption{
Comparison of $\langle\Delta H^2\rangle^{1/2}$
among various 6th order integrators as a function of $\Delta t$.
{\bf Stand} indicates the standard construction scheme of Eq.(\ref{rec}).
{\bf Y1-Y3} indicate Yoshida's construction scheme.
The quenched QCD simulations were done on a $4^4$ lattice at $\beta=5.0$.
}
\end{figure}

\begin{figure}
\centerline{\psfig{figure=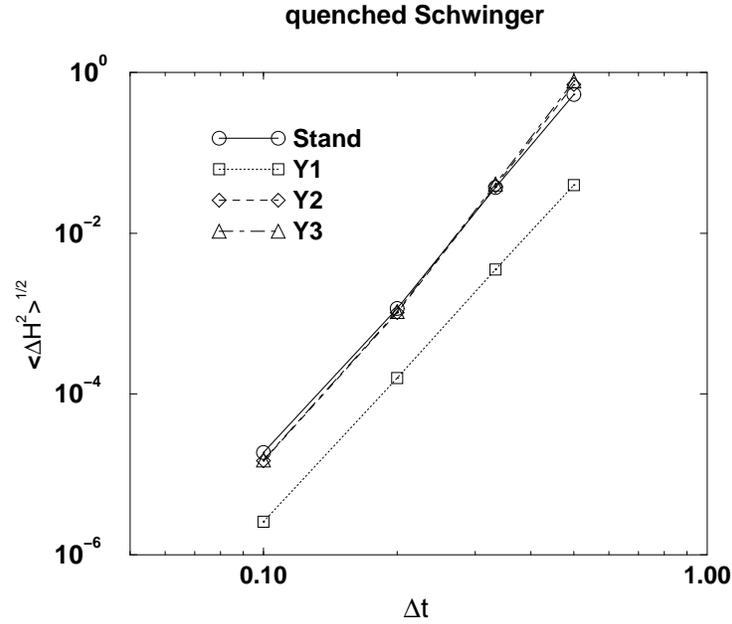,height=8.5cm}}
\caption{
Same as in Fig.4 but for quenched Schwinger simulations
on an $8^2$ lattice at $\beta=1.0$.
}
\end{figure}

\begin{figure}
\centerline{\psfig{figure=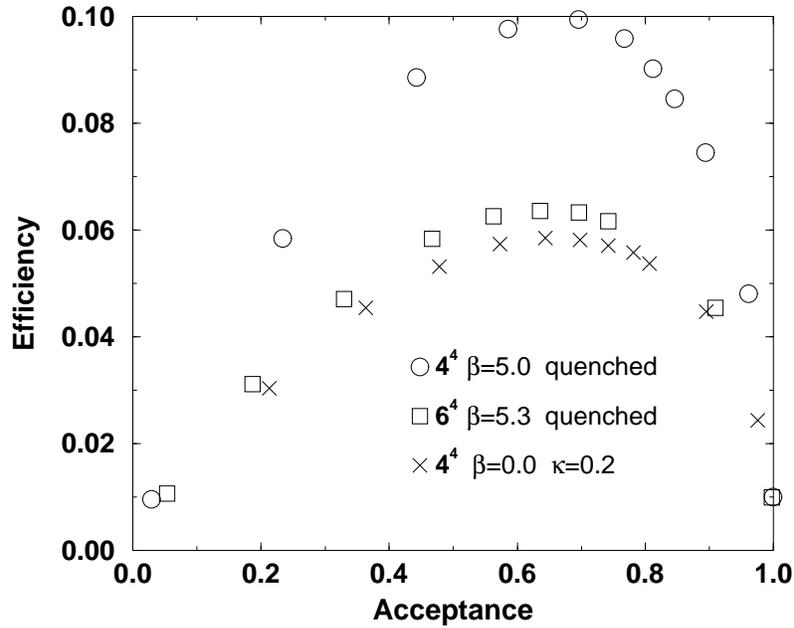,height=8.5cm}}
\caption{
The efficiency $E_{ff}$ of the 2nd order integrator as a function of
the average acceptance.
}
\end{figure}

\begin{figure}
\centerline{\psfig{figure=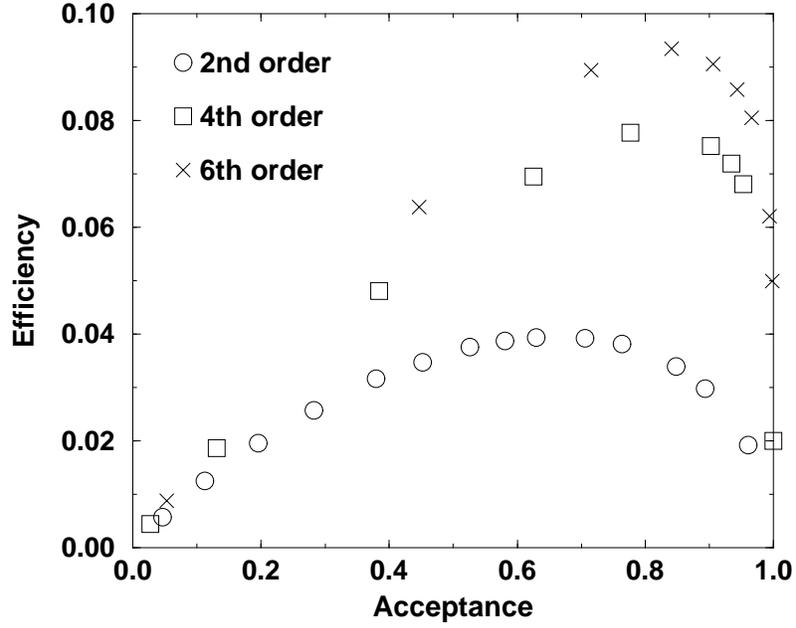,height=8.5cm}}
\caption{
Comparison of the efficiency $E_{ff}$ among the 2nd, 4th and  6th order integrators
for the quenched Schwinger model as a function of the average acceptance.
Simulations were done on a $32^2$ lattice at $\beta=10.0$.
}
\end{figure}

\begin{figure}
\centerline{\psfig{figure=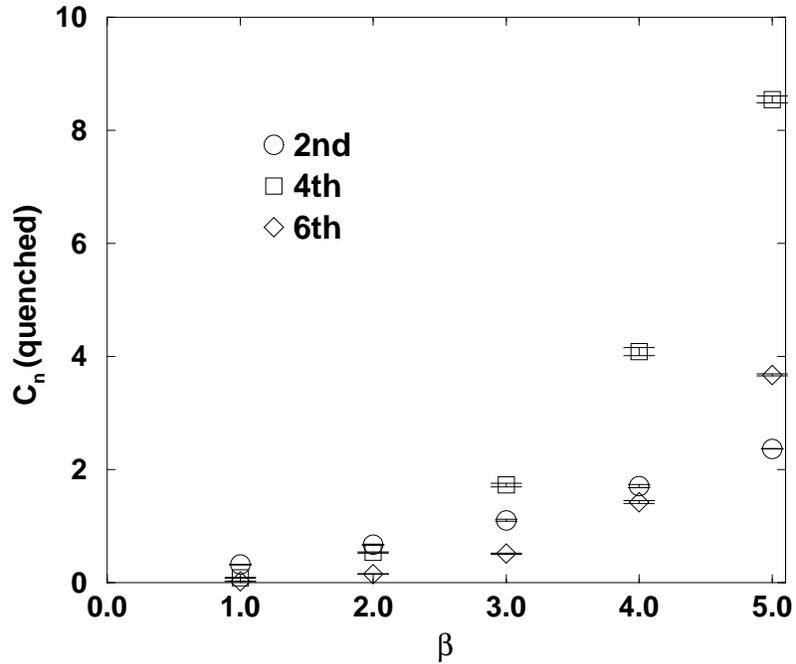,height=9cm}}
\caption{$C_{n}$ for quenched QCD as a function of $\beta$.
}
\end{figure}
\begin{figure}
\centerline{\psfig{figure=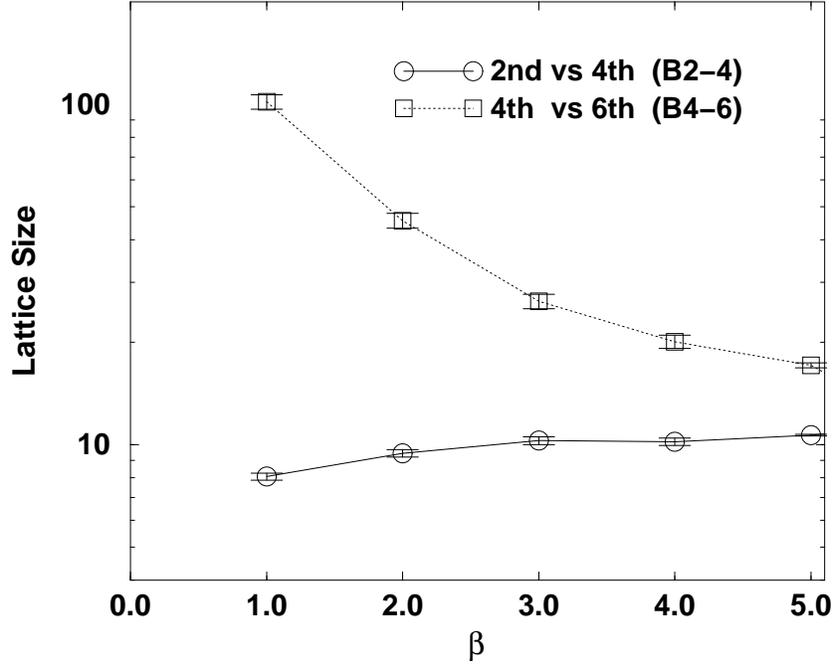,height=9cm}}
\caption{
Results of the boundaries {\bf B2-4} and {\bf B4-6} for quenched QCD.
Lines are shown to guide the eye.
}
\end{figure}

\begin{figure}
\centerline{\psfig{figure=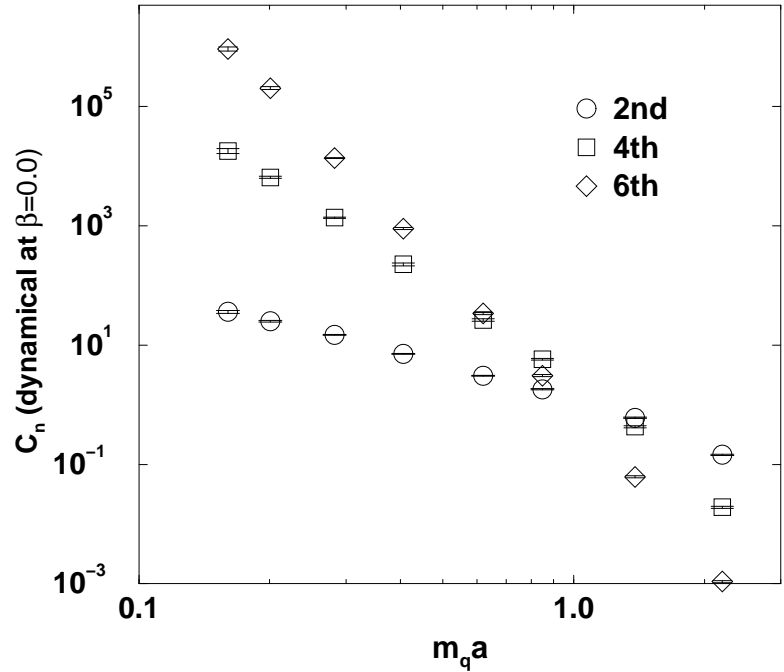,height=9cm}}
\caption{
$C_{n}$ for full QCD as a function of $m_q a$,
where $m_q a =\ln(1+\frac{1}{2}(1/\kappa -1/\kappa_c))$.
}
\end{figure}

\begin{figure}
\centerline{\psfig{figure=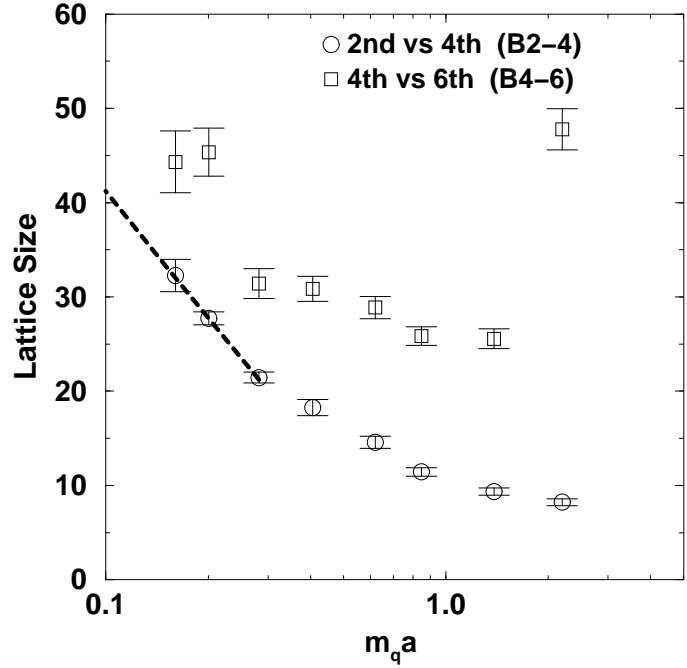,height=9cm}}
\caption{
Results of the boundaries {\bf B2-4} and {\bf B4-6} for full QCD at $\beta=0.0$.
The dashed line is an anticipated boundary of {\bf B2-4}.
}
\end{figure}

\begin{figure}
\centerline{\psfig{figure=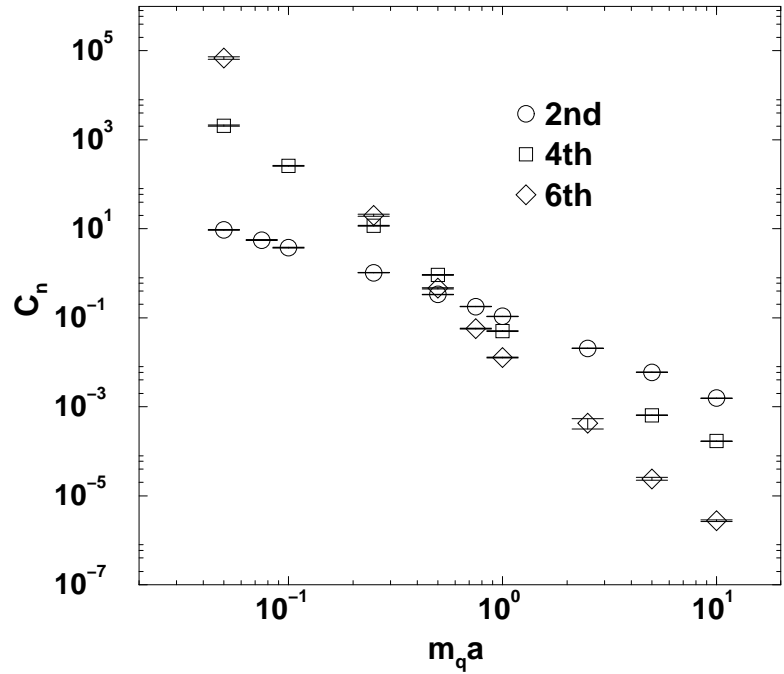,height=9cm}}
\caption{
$C_{n}$ for the Schwinger model with staggered quarks  at $\beta=0.0$.
}
\end{figure}

\begin{figure}
\centerline{\psfig{figure=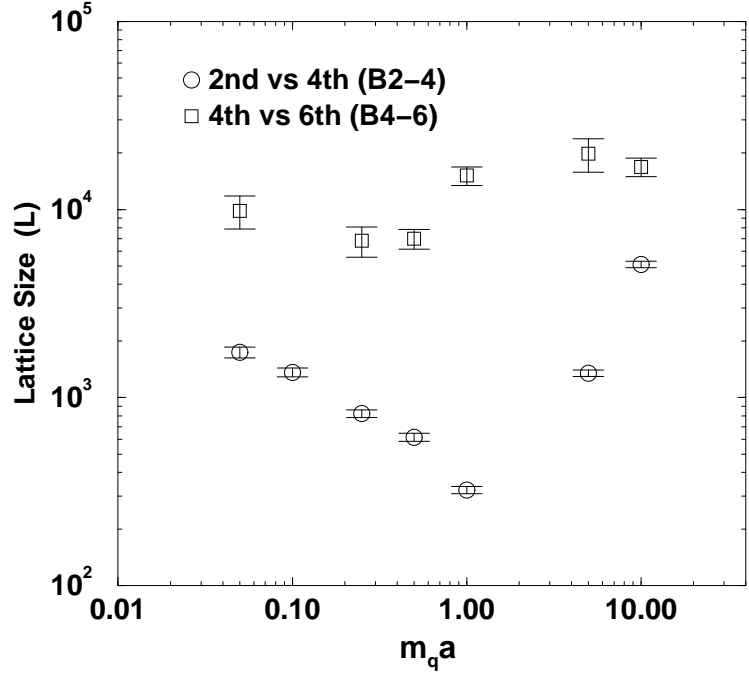,height=9cm}}
\caption{
Results of the boundaries {\bf B2-4} and {\bf B4-6} for Schwinger model 
with staggered quarks  at $\beta=0.0$.
}
\end{figure}

\begin{figure}
\centerline{\psfig{figure=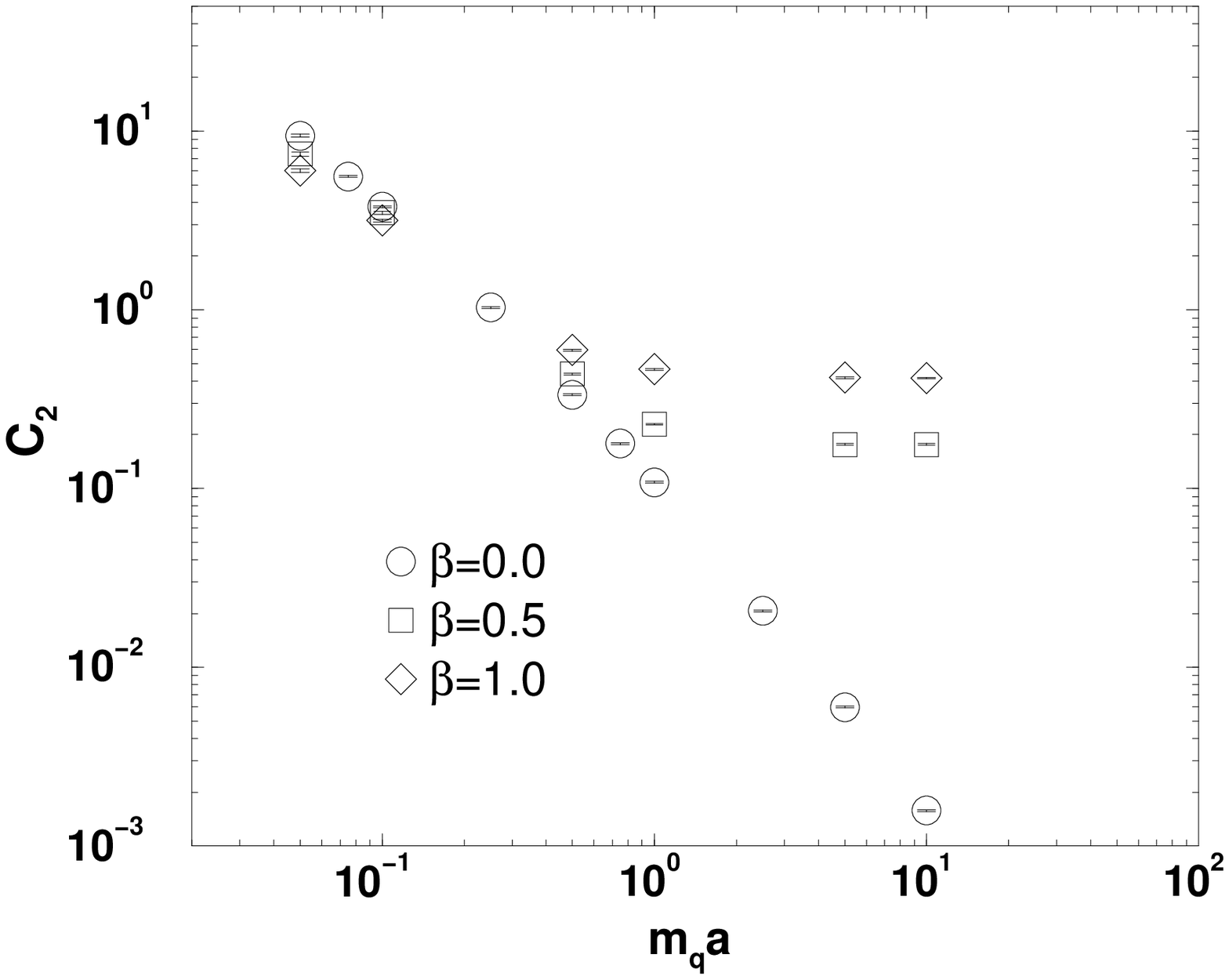,height=9cm}}
\caption{
$C_{2}$ versus $m_qa$ for the Schwinger model with staggered quarks at $\beta=0.0$,
0.5 and 1.0. 
}
\end{figure}


\begin{thebibliography}{99}
\bibitem{HMCA}
S.Duane, A.D.Kennedy, B.J.Pendleton and D.Roweth, Phys. Lett. {\bf B195},
216 (1987);
S.Gottlieb, W.Liu, D.Toussaint, R.L.Renken and R.L.Sugar,
Phys. Rev. D {\bf 35}, 2531 (1987)
\bibitem{SOLVER}
e.g. A.Frommer, Nucl. Phys. B ( Proc. Suppl. ) {\bf 53}, 120 (1997)
\bibitem{Gupta_t}
R.Gupta,G.W.Kilcup and S.R.Sharpe, Phys. Rev. D {\bf 38}, 1278 (1988)
\bibitem{SW}
J.C.Sexton, D.H.Weingarten, Nucl. Phys. {\bf B380}, 665 (1992)
\bibitem{Forcrand}
Ph. de Forcrand, Nucl. Phys. B ( Proc. Suppl. ) {\bf 47}, 228 (1996) 
\bibitem{adaptive}
Ph. de Forcrand and T.Takaishi, Phys. Rev. E {\bf 55}, 3658 (1997)
\bibitem{FULL}
T$\chi$L-Collaboration, Nucl. Phys. B (Proc. Suppl.) {\bf 53}, 222 (1997);
CP-PACS Collaboration, Nucl. Phys. B (Proc. Suppl.) {\bf 73}, 192 (1999) 
\bibitem{WILSON}
K.G.Wilson, Phys. Rev. D {\bf 10}, 2445 (1974);
in New Phenomena in Subnuclear Physics, ed. A.Zichichi, 69 (New York, Plenum, 1975)
\bibitem{Gupta_d}
R.Gupta, A.Patel, C.F.Baillie, G.Guralnik, G.W.Kilcup and S.R.Sharpe, Phys. Rev. D {\bf 40}, 2072 (1989)
\bibitem{4th}
M.Campostrini and P.Rossi, Nucl. Phys. {\bf B329}, 753 (1990)
\bibitem{HOHMC}
M.Creutz and A.Gocksch, Phys. Rev. Lett. {\bf 63}, 9 (1989)
\bibitem{Yoshida}
H.Yoshida, Phys. Lett. {\bf A150}, 262 (1990)
\bibitem{Suzuki}
M.Suzuki, Phys. Lett. {\bf A146}, 319 (1990)
\bibitem{SYMP}
B.Gladman, M.Duncan and J.Candy,
Celestial Mechanics {\bf 52}, 221 (1991)
\bibitem{Accept}
S.Gupta, A.Irb\"ack, F.Karch and B.Petersson,
Phys. Lett. {\bf B242}, 437 (1990)
\bibitem{Creutz}
M.Creutz, Phys. Rev. D {\bf 38}, 1228 (1988) 1228
\end{thebibliography}
\end{document}